\documentstyle[11pt, aaspp4]{article}
\slugcomment{To appear in Astrophysical Journal, Dec 20 issue}

\def\spose#1{\hbox to 0pt{#1\hss}}
\newcommand\lsim{\mathrel{\spose{\lower 3.0pt\hbox{$\mathchar"218$}}
     \raise 2.0pt\hbox{$\mathchar"13C$}}}
\newcommand\gsim{\mathrel{\spose{\lower 3.0pt\hbox{$\mathchar"218$}}
     \raise 2.0pt\hbox{$\mathchar"13E$}}}
\newcommand\msun{{\rm \,M_\odot}}

\begin{document}
\title{N-BODY SIMULATIONS OF COMPACT YOUNG CLUSTERS\\
NEAR THE GALACTIC CENTER}
\author{Sungsoo S. Kim\altaffilmark{1}, Donald F. Figer\altaffilmark{2},
Hyung Mok Lee\altaffilmark{3}, and Mark Morris\altaffilmark{1}}

\altaffiltext{1}{Division of Astronomy, University of California,
Los Angeles, CA, 90095-1562; sskim@astro.ucla.edu; morris@astro.ucla.edu}
\altaffiltext{2}{Space Telescope Science Institute, 3700 San Martin
Drive, Baltimore, MD 21218; figer@stsci.edu}
\altaffiltext{3}{Astronomy Program, SEES, Seoul National University,
Seoul 151-742, Korea; hmlee@astro.snu.ac.kr}

\begin{abstract}
We investigate the dynamical evolution of compact young star clusters (CYCs)
near the Galactic center (GC) using Aarseth's Nbody6 codes.  The relatively
small number of stars in the cluster (5,000--20,000) makes real-number
N-body simulations for these clusters feasible on current workstations.
Using Fokker-Planck (F-P) models, \markcite{KML99}Kim, Morris, \& Lee (1999)
have made a survey of cluster lifetimes for various initial conditions,
and have found that clusters with a mass $\lsim 2 \times 10^4 \msun$
evaporate in $\sim 10$~Myr.  These results were, however,
to be confirmed by N-body simulations because some extreme cluster
conditions, such as strong tidal forces and a large stellar mass
range participating in the dynamical evolution, might violate
assumptions made in F-P models.  Here we find that, in most cases,
the CYC lifetimes of previous F-P calculations are 5--$30 \, \%$
shorter than those from the present N-body simulations.
The comparison of projected number density profiles
and stellar mass functions between N-body simulations and {\it HST}/NICMOS
observations by \markcite{Fe99}Figer et al. (1999) suggests that the
current tidal radius of the Arches cluster is $\sim 1.0$~pc, and the following
parameters for the initial conditions of that cluster: total mass of
$2 \times 10^4 \msun$ and mass function slope for intermediate-to-massive
stars of 1.75 (the Salpeter function has 2.35).
We also find that the lower stellar mass limit, the
presence of primordial binaries, the amount of initial mass segregation,
and the choice of initial density profile (King or Plummer models) do not
significantly affect the dynamical evolution of CYCs. 
\end{abstract}

\keywords{celestial mechanics, stellar dynamics --- Galaxy: center ---
methods: n-body simulations --- galaxies: star clusters}

\section{INTRODUCTION}
\label{sec:intro}

The Arches (G0.121+0.017) and Quintuplet (AFGL 2004) clusters are two
extraordinary star clusters near the Galactic center.  They are very
young ($<5$~Myr), compact ($\lsim1$~pc), and only 20--30 parsecs away from the
Galactic center (GC) in projection, while they appear to be as massive as the
smallest Galactic globular clusters ($\sim 10^4 \msun$;
\markcite{Fe99}Figer et al. 1999).  These compact
young clusters (CYCs) have several interesting dynamical characteristics
that distinguish them from globular clusters: 1) CYCs have
very short dynamical and half-mass relaxation timescales ($t_{dyn} \sim
10^{5-6}$~yr and $t_{rh} \sim 10^{6-7}$~yr, respectively); 2) CYCs are
situated in strong tidal fields (the tidal radius of a $10^4 \msun$
cluster located 30~pc from the Galactic center is 
$\sim 1$~pc); and 3) mass segregation may occur on a timescale shorter
than the lifetimes of the most massive stars, such that those massive
stars play an important role in the dynamical evolution of the cluster
(for more details on timescales and the effects of tidal fields, see
\markcite{KML99}Kim, Morris, \& Lee, 1999; hereafter KML).

The fact that we currently observe only two such young clusters
near the GC (in addition to the central cluster right at the GC) raises a
natural question about the lifetimes of CYCs.
Using anisotropic Fokker-Planck (F-P) models, KML
surveyed lifetimes of CYCs for various
initial mass functions (IMFs), cluster masses ($M$), and Galactocentric
radii ($R_g$), and found that clusters with $M \lsim 2 \times 10^4 \msun$
and $R_g \lsim 100$~pc evaporate in
$\lsim 10$~Myr.\footnote{\markcite{PZe99}Portegies Zwart et al. (1999)
have studied with N-body models the evolution of R136, a compact, young
star cluster in the 30 Doradus region of the Large Magellanic Cloud.
This cluster resembles the CYCs near the GC in many ways, except for the
tidal forces, which are speculated to be much weaker than those for the
CYCs near the GC.  We note that \markcite{PZe00}Portegies Zwart et al.
(2000) are independently working on N-body simulations targeted for
the CYCs near the GC at the moment of the submission of the present paper.}
These unparalleled, short evaporation times ($t_{ev}$; defined here as the
time by which $M$, the cluster mass inside the tidal radius,
becomes 5~\% of its initial value) of CYCs
are first due to short $t_{dyn}$ and $t_{rh}$, and strong tidal forces,
but the mass loss accompanying the evolution of massive stars is also
responsible for shortening $t_{ev}$ of clusters that last longer than
$\sim 3$~Myr.

F-P models are statistical models involving distribution functions.
It is the statistical stability and fast computing time of the F-P
models that the survey-type study by KML required.  \markcite{TPZ98}Takahashi
\& Portegies Zwart (1998) found good agreement between anisotropic
F-P models and N-body simulations for globular clusters by adopting an
``apocenter criterion'' and an appropriate coefficient for the speed
of star removal beyond the tidal radius (KML adopted these as well).
However, some extreme conditions of CYCs may be inconsistent with the
assumptions inherent to F-P models.  As discussed in KML, the conditions
required by F-P models, $t_{dyn} \ll t_{rh}$ and $t_{dyn} \ll t_{se}$
($t_{se}$ is the stellar evolution timescale) may be violated, especially
in the core at certain epochs.  Moreover, the active participation of
a large mass range of stars in the dynamics, which is another peculiarity
of CYCs, is difficult to realize in F-P models that embody a mass
spectrum with a restricted number (usually 10-20) of discrete mass
components.
A greater number of components would better express the mass spectrum,
but then the computation time would become accordingly longer.
On the other hand, too small a number of components would not properly
realize the whole mass spectrum.  By fixing the number of stars
in the most massive components, KML tried to carefully account
for a relatively small number of the most massive stars.  Yet, the
results from such a treatment are to be confirmed by more
realistic models, N-body simulations.

CYCs, estimated to have $\sim 10^{3-4}$ stars, are one of a few classes of
systems for which real-number N-body simulations are feasible on current
workstations.  Among many virtues of N-body simulations, the natural
realization of the mass spectrum and the tidal fields is particularly
beneficial to the study of CYCs.  These benefits allow N-body
simulations to treat mass segregation and evaporation of stars exactly,
thus providing better density profiles and mass spectra as a function of
radius.  These are photometric observables, and are partially available
from the {\it HST}/NICMOS observations of the Arches and Quintuplet
(\markcite{Fe99}Figer et al. 1999; the partial availability is owed to
crowding at the cluster center and to the detection limit at the faint end
of the luminosity function).  The Arches and Quintuplet are estimated
to be only $\sim 2$ and $\sim 4$~Myr old, respectively
(\markcite{Fe99}Figer et al. 1999), but their currently observed structures
must already have deviated from the initial ones due to their
rapid dynamical evolution.  With both real-number N-body simulations and HST
observations of the two clusters in hand, one has a rare chance of deriving
not only the current characteristics of these systems, but their initial
conditions as well.

In the present study, we perform a series of N-body simulations
1) to compare the lifetimes of CYCs with those from F-P models
obtained by KML, 2) to find the initial cluster conditions
that best match the current observations of CYCs,
and 3) to test the effects of initial mass segregation and different initial
density profiles on the dynamical evolution of the CYCs.

\section{SIMULATION MODELS}
\label{sec:models}

For N-body simulations in the present study, we use the most recent
version of Aarseth's N-body codes, Nbody6 (\markcite{A99}Aarseth 1999
and references therein).  We modified the Nbody6 codes to implement the full,
non-truncated tidal forces, and we adopted the prescription for stellar
evolution used in KML for consistency (see KML for details on the stellar
evolution).  The upper stellar mass limit, $m_u$, is $150 \msun$, and
the lower mass limit, $m_l$, is $0.1 \msun$ or $1 \msun$.
We set the cluster to initially fill the tidal radius, and remove
stars outside 5 times the tidal radius.

For initial conditions of the cluster and the Galactic tidal field, we
also follow prescriptions of KML:  We adopt single-mass King models
(with a King parameter $W_0 =4$) and single power-law mass functions;
The Galactic mass $M_g$ inside a Galactocentric radius $R_g$ is given by
\begin{equation}
\label{bulgemass}
	M_g = 2 \times 10^8 \msun \, \left ( {R_g \over
		{\rm 30 \, pc}} \right )^{1.2}
\end{equation}
(from \markcite{GT87}Genzel \& Townes, 1987).
Then the tidal radius of the cluster, $R_t$, becomes
\begin{eqnarray}
	R_t & =      & \left ( {M \over 2 M_g} \right )^{1/3} R_g \nonumber\\
	    & \simeq & 5.3 \times 10^{-3} \, {\rm pc} \, \left ( { M \over
			{\rm M_\odot}} \right )^{1/3} \left ( {R_g \over
			{\rm pc}} \right )^{0.6}.
\end{eqnarray}
The equations of motions are integrated in the rotating frame, and tidal
forces are calculated from the Galactic potential corresponding to
equation~(\ref{bulgemass}).

\section{N-BODY VS. FOKKER-PLANCK}
\label{sec:nbody2fp}

In order to compare $t_{ev}$ of N-body and F-P calculations, we performed
N-body simulations of
several representative models in KML that have number of stars, $N$,
smaller than $\sim 15,000$.  This criterion excludes $m_l = 0.1 \msun$
models with the exponent of the power-law IMF, $\alpha$ (defined as in
$dN \propto m^{-\alpha} dm$), larger than or equal to 2.

The lifetimes of 9 N-body simulations performed in the present
study are shown in Table~\ref{table:nbody2fp} along with those
from F-P calculations of KML.\footnote{The lifetimes from N-body simulations
are subject to the statistical fluctuation which may be significant for
models with $N$ smaller than a few thousands.  We find that our models
have fluctuations from one run to the next
mostly smaller than 10~\%.  Such relatively small
fluctuations are due to our definition of $t_{ev}$ (see \S~\ref{sec:intro}),
which avoids the late phase of the evolution that suffers most from statistical
fluctuations.}
This set of 9 models represent
different $M$ (models 1 \& 5), different $\alpha$ (models 1, 8, 13, \& 23),
different $m_l$ (models 3 \& 8; models 13 \& 15), and different
$R_g$ (models 13, 43, \& 44).  We find that N-body calculations always give
longer $t_{ev}$ than F-P calculations, and that the fractional difference
of $t_{ev}$ between two calculations, $\Delta' t_{ev}$
($\equiv$~[Nbody-FP]/Nbody), does not show any particular correlation
with $N$.  The biggest $\Delta' t_{ev}$, 59~\%, is
for model 43, for which the IMF is quite flat ($\alpha = 1.75$) and
the cluster evaporates before significant stellar evolution starts.
This implies that
$\Delta' t_{ev}$ is larger for smaller $\alpha$ (compare to model 5, which has
as short $t_{ev}$ as model 43 but $\Delta' t_{ev}$ of only 8~\%), but that
the stellar evolution has the effect of diminishing $\Delta' t_{ev}$
(cf. model 44, which also has $\alpha=1.75$ but $\Delta' t_{ev}$ of only
16~\%).  We performed the N-body and F-P calculations for model 13
without stellar evolution, and find that the new $t_{ev}$'s
are 4.5 and 2.5~Myr, respectively (model 13 is our baseline model
because we later find that it best matches the observations;
see \S~\ref{sec:init}).  This indicates that the speed of
the pure (i.e.,  no stellar evolution) dynamical evolution of F-P models
by KML is considerably overestimated.

We attribute shorter $t_{ev}$'s of F-P calculations to the underestimated
choice of $N_u$ (the number of stars in the most massive bin; the same
as $N_{15}$ when the number of mass bins, $N_{\rm bin}$, is 15 as in KML)
by KML.  The mass bins of F-P models by KML were chosen to make $N_u$ constant,
so that the significance of the most massive bins of models with
different IMFs would be the same.  The larger $N_u$, the smaller
the characteristic mass of the most massive bin, thus the
smaller the effective mass range (the lower and upper mass limits
are not a function of $N_u$).
Since a larger effective
mass range gives faster mass segregation and thus faster ejection
of lighter stars, a smaller $N_u$ results in smaller $t_{ev}$.
For $N_{\rm bin}$ cannot be infinitely large for a practical reason,
F-P models have difficulties with dealing with a very large mass range,
and F-P results may depend on the way of binning the mass with
a limited number of components (bins).
The choice of $N_u=50$ in KML was arbitrary, and one may tune
the F-P $t_{ev}$ values to N-body $t_{ev}$ values by controlling
this parameter.  We find that a new F-P calculation for model
13 with no stellar evolution and $N_u=250-300$ gives a $t_{ev}$ value
that matches the N-body results without stellar evolution, 4.5~Myr.
In conclusion, the choice of $N_u=50$ in KML was too small, thus KML
underestimated $t_{ev}$, but such underestimation was not significant
for most cases because the amount of mass loss from stellar evolution after
$\sim 2$~Myr accelerates the evaporation of clusters.

\begin{figure}
\centerline{\epsfxsize=8.8cm\epsfbox{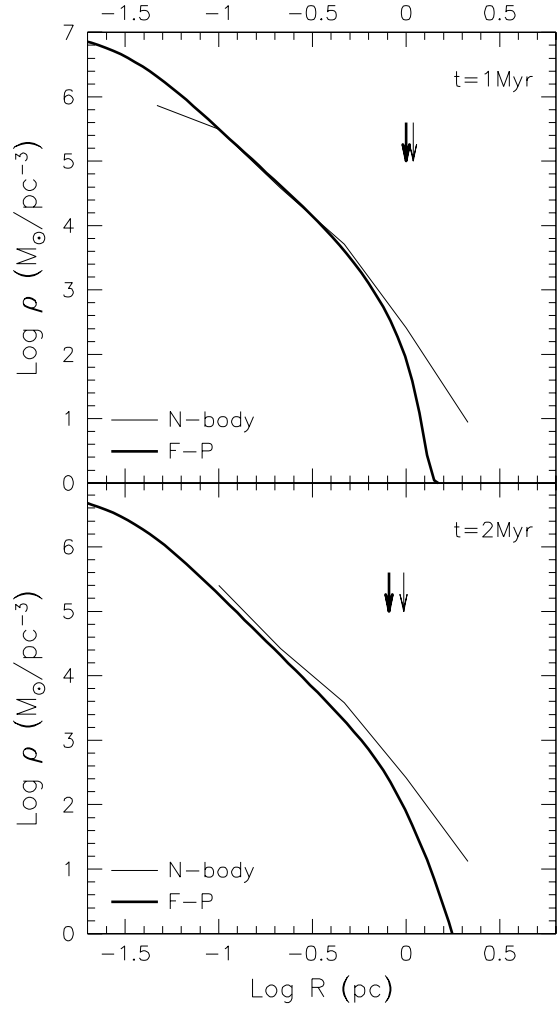}}
\caption
{\label{fig:rho}Volume density profiles of N-body ({\it thin lines})
and F-P ({\it thick lines}) calculations for model 13 at $t=1$~Myr ({\it upper
panel}) and 2~Myr ({\it lower panel}).  The locations of tidal radii are
marked with arrows.
$N_u=250$ and $N_{\rm bin}=10$ were used for the F-P calculation.
}
\end{figure}

\begin{figure}
\centerline{\epsfxsize=8.8cm\epsfbox{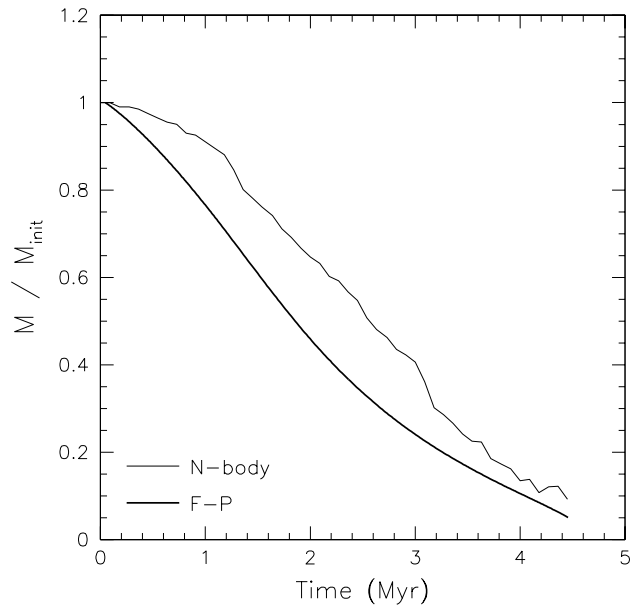}}
\caption
{\label{fig:mtot}
Evolution of cluster mass inside the tidal radius, $M$, of N-body ({\it thin
lines}) and F-P ({\it thick lines}) calculations for model 13.  $M$'s are
normalized to their initial values.  Stellar evolution is not included in
these calculations, and $N_u=250$ and $N_{\rm bin}=10$ were used for the
F-P calculation.
}
\end{figure}

Figure~\ref{fig:rho} compares the volume density profiles of N-body
and F-P calculations for model 13 at 1 and 2~Myr.  For this F-P
calculations, $N_u =250$ and $N_{\rm bin}=10$ (instead of 15 as in KML)
were used.  For $N_u =250$, $N_{\rm bin}$ larger than 10 would give
negative $\beta$ values (see KML for definition of $\beta$; a negative
$\beta$ gives smaller numbers of stars for lighter mass bins).
At $t=1$~Myr, the two density profiles differ only near and outside
$R_t$ (the small deviation at the core is probably due to small
number statistics), but the F-P calculation significantly underestimates the
density outside $R_t$.  At $t=2$~Myr, the deviation starts at smaller
radii, and the power-law slope of the F-P density profile is steeper than
the N-body slope by 0.3 to 0.4 inside $R_t$.
Such a difference makes the cluster mass inside the tidal radius of the F-P
calculation always somewhat smaller than that of the N-body calculation.
Figure~\ref{fig:mtot} compares the evolution of cluster mass inside the
tidal radius, $M$, of N-body and F-P calculations for model 13.  Stellar
evolution was not included in these calculations in order to see the effect
of tidal fields only, and $N_u=250$ and $N_{\rm bin}=10$ were used for the
F-P calculation.  The figure shows that the mass loss rate from the cluster
differs especially during the early phase of evolution.
We also find that
the $\alpha$ value of the F-P calculation decreases faster with time,
having 0.1 to 0.3 smaller (shallower mass function) values, than those
of the N-body simulation at most times.  N-body models are more realistic
than F-P models, whose observables are here found to have non-negligible
discrepancies from N-body models; therefore for comparison with the observed
structures of the CYCs, N-body simulations should be preferred over
F-P models.

\section{INITIAL CONDITIONS OF THE ARCHES}
\label{sec:init}

\subsection{Comparison with Observations}
\label{subsec:comp_obs}

In spite of the very young ages of the CYCs, the observed properties of
CYCs should differ from their initial status because of the cluster's rapid
dynamical evolution.  Thus one should compare numerical simulations
of cluster evolution with observations to infer the initial conditions
of the cluster.  As discussed in \S~\ref{sec:nbody2fp}, N-body
simulations provide the most accurate information on the dependence of
the stellar distribution on stellar mass.  In this section, we use our N-body
simulations to find initial conditions of the Arches cluster from
comparisons with {\it HST}/NICMOS photometric observations by
\markcite{Fe99}Figer et al. (1999).  The Quintuplet cluster has a largely
dispersed distribution (probably because it is in the final disruption
phase), so dynamical information from its image is very limited.

Dynamical information from the photometric image of a cluster can be
found from the surface density profiles and the mass distribution functions
as a function of radius.  When the photometry is complete down to
the faintest stars, the overall stellar mass function, total cluster
mass, and degree of mass segregation are obtained from the above two
observables.  However, the {\it HST} observations of the Arches were
limited by crowding in the core and by the background confusion in the
outer regions.  Using the mass-K magnitude relation of
\markcite{Me94}Meynet et al. (1994) and adopting an extinction
at K band of 3.1, \markcite{Fe99}Figer et al. (1999) estimate that their
{\it HST} Arches photometry is complete down to $m \simeq 20 \msun$
for $r > 3 \arcsec$, or 0.12~pc, and to $m \simeq 8 \msun$
for $r > 5.\arcsec 25$, or 0.2~pc (the distance to the GC is
assumed to be 8~kpc in the present study), and that the background
confusion limit lies between 3 and $5\msun$.  For this reason,
we only use stars having $m>20 \msun$ (${\rm F205W}<15$~mag, where
F205W is the undereddened, apparent Vega magnitude of the NICMOS F205W
filter) for the surface number density profile, and $m>8 \msun$
(${\rm F205W} <17$~mag) for estimating the mass spectrum.  The number
of stars heavier than $8 \msun$ outside its crowding limit, 0.2~pc,
is 232, which is not enough to give reliable information on the mass
spectrum as a function of radius.  Therefore we adopt the simplest
way of measuring the mass function and the mass segregation:  we count
the number of stars in two mass bins ($8 \leq m/{\rm M_\odot} < 20$ and
$20 \leq m/{\rm M_\odot}$), each in two radius bins ($0.2 \leq r/{\rm pc}
<0.4$ and $0.4 \leq r/{\rm pc} <0.8$).  In principle, one can find the
initial conditions (or several sets of initial conditions) of a cluster
by comparing these number counts and the density profiles of numerical
simulations and observations.

\begin{figure}
\centerline{\epsfxsize=8.8cm\epsfbox{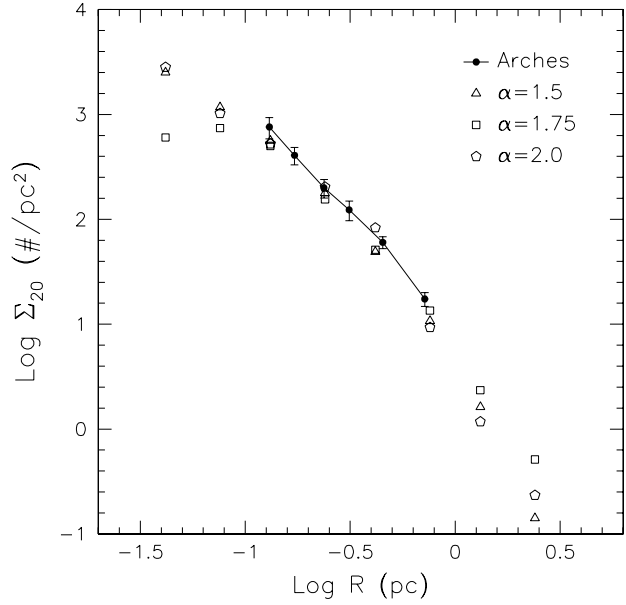}}
\caption
{\label{fig:prof_alpha}Surface number density profiles of stars heavier
than $20 \msun$, $\Sigma_{20}$, from our N-body models 21 ($\alpha = 1.5$;
{\it triangles}), 13 ($\alpha = 1.75$; {\it squares}), \& 22 ($\alpha = 2.0$;
{\it pentagons}), and NICMOS observations of the Arches by
Figer et al. (1999; {\it solid circles connected with a line}).
N-body results shown are the ones
at $t=2$~Myr for models 13 \& 22, and $t=1$~Myr for model 21 (model 21 shows
a worse fit for the core region at $t=2$~Myr).  The tidal radii of N-body
models 21, 13, \& 22 at the epoch shown here are 1.0~pc, 1.0~pc, \& 1.2~pc,
respectively.  The observations are limited by crowding at the core and
by background confusion near and outside the tidal radius.
1-$\sigma$ Poisson errors for observations are indicated by vertical bars.
}
\end{figure}

\begin{figure}
\centerline{\epsfxsize=17cm\epsfbox{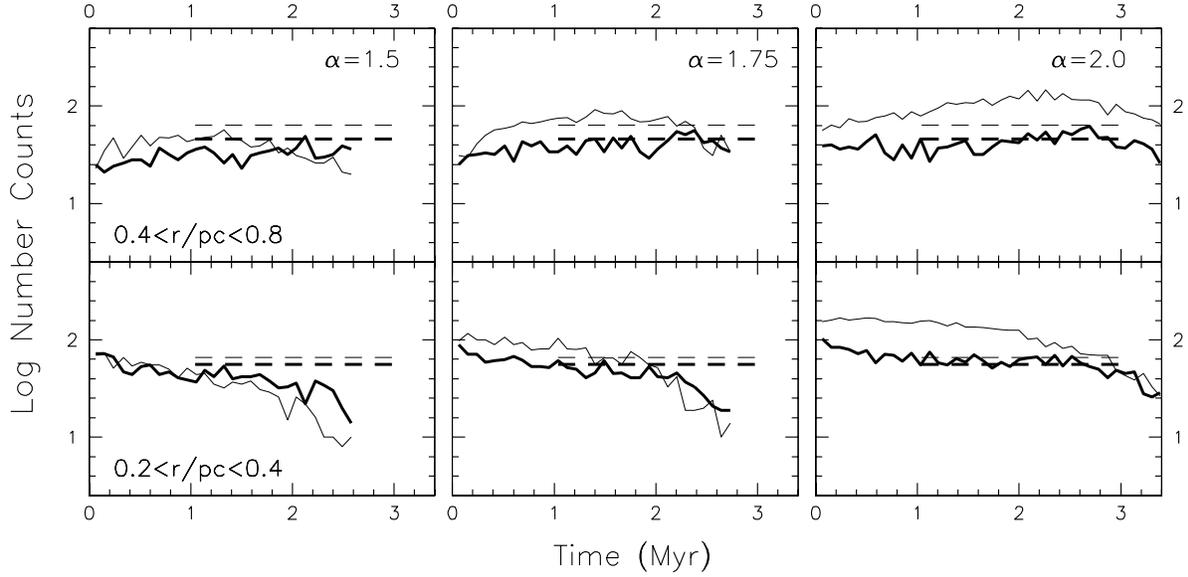}}
\caption
{\label{fig:count_alpha}Evolution of number counts of stars with $8 \le
m/{\rm M_\odot} \le 20$ ({\it thin lines}) and stars with $20 \le m/
{\rm M_\odot}$ ({\it thick lines}) in the $0.2 < r/{\rm pc} <0.4$ region
({\it lower panels}) and the $0.4 < r/{\rm pc} <0.8$ region ({\it upper
panels}), from our N-body models 21 ($\alpha = 1.5$; {\it solid lines in
left panels}), 13 ($\alpha = 1.75$; {\it solid lines in middle panels}),
\& 22 ($\alpha = 2.0$; {\it solid lines in right panels}), and NICMOS
observations of the Arches by Figer et al. (1999; {\it dashed lines}).
The dashed lines cover the estimated age of the Arches cluster, $2 \pm 1$~Myr.
}
\end{figure}

We start with IMFs that give the observed number of massive stars,
which, for the Arches cluster, is expected to be very close to its
initial value (the estimated
age of the Arches, $\sim 2$~Myr, is before significant stellar evolution).
The observations require the IMF to have $N$($>30 \msun$) of 150
($30 \msun$ stars are complete down to $r \simeq 0.06$~pc).  The surface
number density profiles for $M \geq 20\msun$ ($\Sigma_{20}$) at 1 or 2~Myr,
and the number count evolutions of N-body simulations of models 13, 21,
and 22 are plotted in Figures~\ref{fig:prof_alpha} and \ref{fig:count_alpha}
along with observations (the count numbers from observations are given
in Table~\ref{table:obscount}).
These three models have $N$($>30 \msun$) values of about 150
but different $\alpha$ values (see Table \ref{table:init}).  $\Sigma_{20}$
plots in Figure~\ref{fig:prof_alpha} show that all three models agree
well with the observation, which implies that our choice of the bulge
mean density at the location of the cluster is appropriate because the mean
bulge density determines the tidal radius of the cluster (this does
not necessarily prove that our choices of the bulge mass profile,
eq.~[\ref{bulgemass}], and $R_g$ for these models, 30~pc, are correct
since the bulge mean density is a combination of two
parameters; see below).  The $R_t$ values of our N-body models shown
in Figure~\ref{fig:prof_alpha} are 1.0--1.2~pc.
Figure~\ref{fig:count_alpha} shows that the
observed number counts best agree with model 13 ($\alpha=1.75$) at
time $t=1$--2~Myr.  Model 21 ($\alpha=1.5$) gives slightly smaller counts
than observations, but considering the uncertainty in the mass-magnitude
relation, we cannot exclude the possibility that model 21 is
applicable.  On the other hand, model 22 ($\alpha = 2$) significantly
over-predicts the number of stars in lighter mass bins, except at
$t \gsim 3$~Myr, by which time the cluster would be in a disruption
phase.  We conclude that the best $\alpha$ value suggested by
these comparisons to observations is 1.75.

This finding may be, in fact, limited to the high mass regime only, where
the comparison is actually made.  However, for the above best-fit mass
function, the stars with $m \ge 8 \, {\rm M_\odot}$ already consitute
more than 70~\% of the total mass, thus the above $\alpha$ value may be
considered representative.

The projected distance from the GC to the Arches is $\sim 20$~pc, but
the true distance is unknown.  Furthermore, the bulge mass distribution
at a few tens of parsecs from the GC is uncertain.  While the
infrared light distribution observed by \markcite{BN68}Becklin \&
Neugebauer (1968) implies a density profile $\propto R_g^{-1.8}$
(eq.~[\ref{bulgemass}] exhibits this profile), the radial velocity
observations in radio frequencies for the same region suggest a shallower
density drop ($\propto R_g^{-1.5}$) between 30 and 100~pc, leading to
a steeper enclosed-mass increase:
\begin{equation}
\label{bulgemass2}
	M_g \simeq 8 \times 10^7 \msun \left ( {R_g \over {\rm 30 \, pc}}
		     \right )^{1.5}
\end{equation}
(\markcite{LHW92}Lindqvist, Habing, \& Winnberg 1992).  The tidal environment of the
cluster is mostly determined by the mean density, $M_g / R_g^3$.
We performed the best-fit model found above ($\alpha=1.75$; model 13)
with different tidal environments: models located at $R_g =20$~pc
(model 19) and 50~pc (model 14) with our standard bulge mass distribution,
equation~(\ref{bulgemass}).  The latter is equivalent to the case of
$R_g=30$~pc with a bulge mass distribution of equation~(\ref{bulgemass2}).
We find that $\Sigma_{20}$ of models 13, 14, and 19 can be all nicely
fit to the observed $\Sigma_{20}$ at $t=1$--2~Myr, and that the number
counts of models 14 and 19 showed worse agreement to observations than
model 13, but the agreement was still at the acceptable level.  We
conclude that our comparisons do not rule out the possibility of
$R_g=20$~pc and 50~pc, but favor the $R_g=30$~pc case.

Model 15 has the same initial conditions as our best-fit model,
model 13, except $m_l = 0.1 \msun$ instead of $1 \msun$.
We find that $t_{ev}$
of model 15 is only $\sim 10 \, \%$ longer than that of model 13
(see Table~\ref{table:init}).  This confirms the finding by KML
that $t_{ev}$ does not sensitively depend on $m_l$ when $\alpha \lsim 2$.
This phenomenon is caused by two factors: 1) for a model with
$\alpha = 1.75$ and $m_u = 150 \msun$, the mass between
$0.1 \msun$ and $1 \msun$ constitutes only $15 \, \%$ of the
total mass, and 2) the lightest stars are rapidly ejected from the
cluster during the early phases due to the large $m_u/m_l$ ratio.
However, the number counts of model 15 agree well with observations
only after $t=2.4$~Myr, by which time the Arches cluster would be in the
disruption phase and would have a structure as dispersed as the Quintuplet.
Thus the observations more support our model with $m_l = 1 \msun$
than $0.1 \msun$.

One of the most commonly used IMFs for the Galactic disk,
the Scalo mass function (\markcite{S86}Scalo 1986), has $\alpha=2.7$
for $m \ge 1\msun$, implying that the best-fit $\alpha$ value obtained
here for the Arches cluster is considerably flatter than that for the disk.
This fact, together with $m_l = 1 \msun$ being favored over $m_l =
0.1 \msun$, appears to support the arguments by \markcite{M93}Morris
(1993) that the non-standard star formation environment near the GC may
lead to an IMF skewed toward relatively massive stars and having
an elevated lower mass cutoff.

\subsection{Primordial Binaries}
\label{subsec:binaries}

So far, we have not considered primordial binaries.  Camera 2 of the
{\it HST}/NICMOS instrument (used by \markcite{Fe99}Figer et al. 1999)
has an angular resolution of about 1~pixel size of the detector
($0.075 \arcsec$), which is $\sim 600$~AU at the distance of the GC.
Thus binary systems with a semi-major axis smaller than few hundred AU
are not resolved, and the presence of a significant number of primordial
binaries in the cluster may affect our number count analysis above
by decreasing number counts and/or by moving a primary star to a more
massive mass bin (the number of dynamical binaries, i.e., binaries
formed through close encounters, at a given moment is only a few,
if any, for the whole cluster lifetime).  To see the effects of
primordial binaries on the number counts, we performed two simulations
of model 13 with primordial binary fractions, $f_{bin}$, of 25~\%
and 50~\%.  The fraction is defined as
\begin{equation}
\label{binfrac}
	f_{bin} \equiv {N_{bin} \over N_{bin} + N_{sing}},
\end{equation}
where $N_{bin}$ and $N_{sing}$ are the numbers of binary systems and
single stars, respectively.  Thus the percentages of stars in binary
systems for $f_{bin}=25$~\% and 50~\% are 40~\% and 67~\%, respectively.
These relatively `moderate' $f_{bin}$ values, compared to $>50$~\% used in
N-body simulations for open clusters (\markcite{KPM99}Kroupa, Petr, \&
McCaughrean 1999, for example) may be justified by an argument by
\markcite{DS94}Durisen \& Sterzik (1994) that binary formation from
fragmentation of collapsing and rotating clouds or from a gravitational
instability of massive protostellar disks is more likely in low-temperature
clouds (the central molecular zone in the inner few hundred pc of the
Galactic bulge has significantly elevated temperatures, $\sim 70$~K).
The initial companion-mass-ratio distribution is obtained by random
pairing of stars, and the initial eccentricity distribution is assumed
to be thermally relaxed.  For the initial period distribution,
we adopt equation~(8) of \markcite{K95a}Kroupa (1995a), which approximates
the distribution of binary systems in the Galactic disk.  This initial
distribution is evolved prior to the start of the N-body integration
to account for the ``pre-main-sequence eigenevolution" (the evolution
in orbital parameters due to internal processes such as tidal circularization
during the pre-main-sequence; see \markcite{K95a}Kroupa 1995a for details).

We find that the $t_{ev}$ values of $f_{bin}=25$~\% and 50~\% models
are slightly (10--20~\%) longer than that of $f_{bin}=0$~\%
model.  This insensitivity of $t_{ev}$ on $f_{bin}$
was noted by \markcite{MH94}McMillan \& Hut (1994), and \markcite{K95b}Kroupa
(1995b).  \markcite{KPM99}Kroupa et al. (1999) find that
their model with $f_{bin}=100$~\% evolves slightly more slowly than
that with $f_{bin}=60$~\% and interpret it due to cooling by disruption
of wide binaries.  The slightly longer $t_{ev}$'s of our models with primordial
binaries may be interpreted in the same way, but the reduced number of
effective point sources due to binarity may be another possible explanation
(the rate of relaxation is proportional to $Nm^2$, where $m$ is the stellar
mass; when companions are picked out of a pool with a large mass range,
the binarity does not significantly increase the $m$ of an `effective point
mass', while it considerably decreases $N$ of the effective point masses).

\begin{figure}
\centerline{\epsfxsize=12.cm\epsfbox{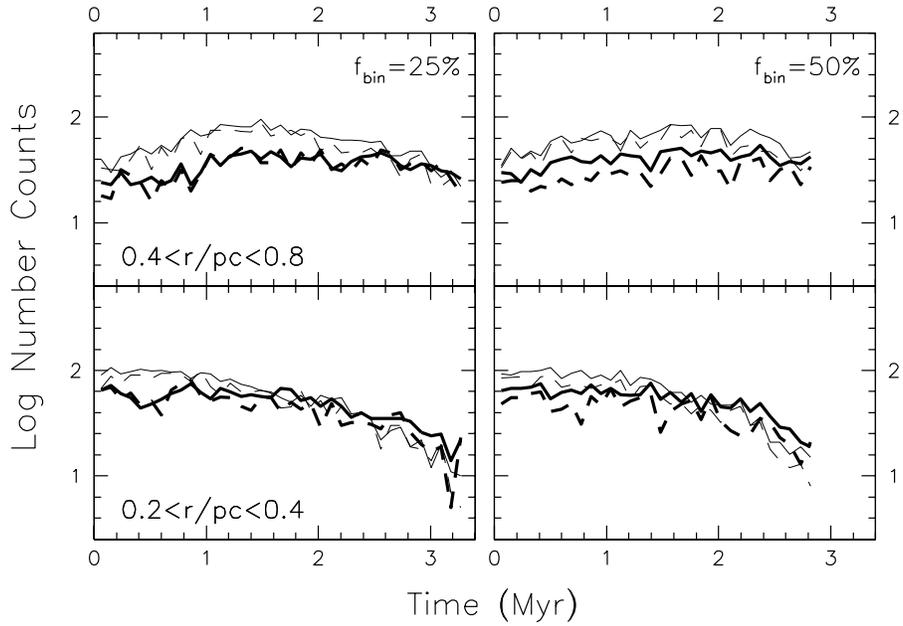}}
\caption
{\label{fig:count_bin}Evolution of number counts of stars with $8 \le
m/{\rm M_\odot} \le 20$ ({\it thin lines}) and stars with $20 \le m/
{\rm M_\odot}$ ({\it thick lines}) in the $0.2 < r/{\rm pc} <0.4$ region
({\it lower panels}) and the $0.4 < r/{\rm pc} <0.8$ region ({\it upper
panels}), from our N-body model 13 with $f_{bin}=25$~\% ({\it dashed lines
in left panels}) and model 13 with $f_{bin}=50$~\% ({\it dashed lines in
right panels}).  The angular resolution of the {\it HST}/NICMOS camera 2,
$\sim 0.075 \arcsec$, is considered in counting apparent numbers and
estimating apparent masses (see text for details).
Model 13 with $f_{bin}=0$~\% ({\it solid lines in both panels})
is also plotted for comparison.
}
\end{figure}

The number counts with a consideration of the angular resolution of the
camera for $f_{bin}=25$~\% and 50~\% models are compared to those of
the $f_{bin}=0$~\% model in Figure~\ref{fig:count_bin}.
Here, the mass of a binary system with a semi-major axis smaller
than the angular resolution is obtained from the total luminosity of the
binary system, and the mass-luminosity relation adopted in
\markcite{Fe99}Figer et al. (1999) is used for this calculation.
While the number counts of the $f_{bin}=25$~\% model are only slightly
smaller than the $f_{bin}=0$~\% model, those of the $f_{bin}=50$~\% model are
10--40~\% smaller.  However, these relatively smaller number
counts of models with primordial binaries still agree with observed
number counts at an acceptable level.
Furthermore, the difference of number counts between
the $f_{bin}=0$~\% and 50~\% models is larger for the more massive bin,
implying that the initial $\alpha$ should be smaller than that of the model
tried here (1.75) in order for the $f_{bin}=50$~\% model to better match
the observations.  In conclusion, the presence of primordial binaries
will not significantly change our findings above on the best-fit
initial conditions of the Arches cluster, and our results indicate that
the $\alpha$ value found above for the case with no primordial binaries, 1.75,
is an upper limit.

\section{INITIAL CLUSTER STRUCTURE}
\label{sec:structure}

\subsection{Initial Mass Segregation}
\label{subsec:segregation}

The prior sections of the present study and KML assumed, for simplicity,
the same initial density profiles for all stellar masses,
i.e., no initial mass segregation.  This assumption is, however,
not based on any observational or theoretical evidence.
Interactions between stars during the
star formation process determine the degree of initial mass
segregation, which is therefore an important piece of information
for the theory of cluster formation.  However, different models lead to
totally different predictions for initial segregation. The model by
\markcite{PP92}Podsiadlowski \& Price (1992), in which favorable
stellar masses are determined by the ratio of the timescales for
protostellar collisions and of gas infall onto the protostars,
predicts more massive stars at larger radii. On the other hand,
segregation of more massive stars in the core is predicted in some models
such as the one by \markcite{ML96}Murray \& Lin (1996), where encounters
between cloudlets increase the protostellar masses, and the one by
\markcite{Be97}Bonnell et al. (1997), where the deeper potential
in the core causes stars there to accrete more circumstellar material.
Here we attempt both cases, one with heavy stars initially more prevalent
at the core, the other with heavy stars more in the envelope.

\begin{figure}
\centerline{\epsfxsize=8.8cm\epsfbox{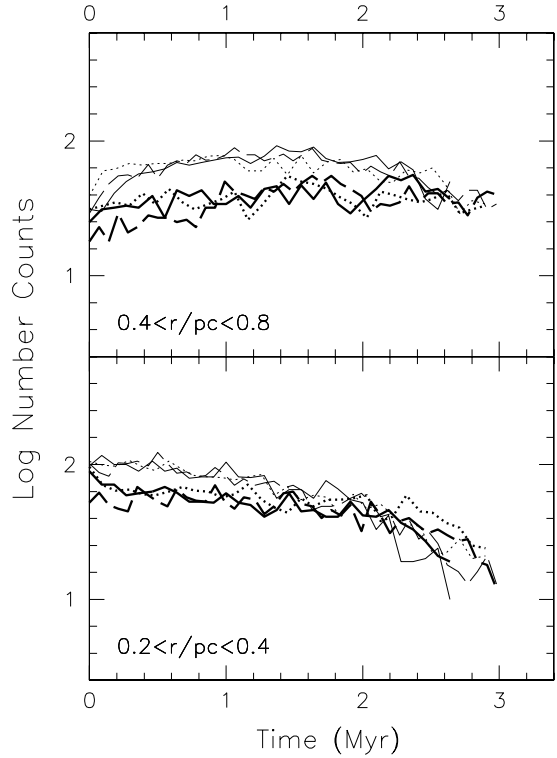}}
\caption
{\label{fig:count_seg}Evolution of number counts of stars with $8 \le
m/{\rm M_\odot} \le 20$ ({\it thin lines}) \& stars with $20 \le m/
{\rm M_\odot}$ ({\it thick lines}) in the $0.2 < r/{\rm pc} <0.4$ region
({\it lower panel}) and the $0.4 < r/{\rm pc} <0.8$ region ({\it upper
panel}), from our N-body models 13 (no initial segregation; {\it solid lines}),
31 (initially more heavy stars in the evelope; {\it dashed lines}), \& 32
(initially more heavy stars in the core; {\it dotted lines}).
}
\end{figure}

The initial density and velocity profiles of model 13 are modified
to include the initial mass segregation: model 31 initially has
a King profile with $W_0=2$ for the lightest stars and $W_0=6$
for the heaviest stars (more heavy stars in the envelope) while
model 32 has the opposite $W_0$ values (more heavy stars at the core).
The intermediate-mass stars have interpolated $W_0$ values depending
on the logarithm of their masses.  We find that models 13, 31, \& 32
show very similar $\Sigma_{20}$ profiles at 2~Myr.  In spite of
different initial conditions, the number counts of these three models also
exhibit similar evolution after $t=1$~Myr (see Fig.~\ref{fig:count_seg}).
It appears that the relaxation processes are rapid enough to erase the
memory of the initial mass segregation in less than 1~Myr.  Thus unless
the initial segregation is more severe than models tested here, the
best-fit initial conditions found in \S~\ref{sec:init} are robust
against the initial segregation.

\subsection{King vs. Plummer Models}
\label{subsec:plummer}

KML showed that the global evolution (such as $M$ \& $R_t$) of
clusters initially having King profiles of $W_0=1$--7 does not depend
on $W_0$ values.  This is because King profiles with different
$W_0$ values mainly differ at the core while the global evolution
of the cluster is mostly determined by the properties in the cluster
envelope.  Here we make two N-body simulations initially having
Plummer models to see the effects of initial conditions different
from King models.  Plummer models have a density profile of
\begin{equation}
\label{plummer}
	\rho \propto (R^2 + R_c^2)^{-5/2},
\end{equation}
which has no density drop analogous to tidal cutoff in King models
($R_c$ is the core radius).
Thus the density profiles of Plummer models have shallower decrease
than King models near $R_t$, and have to be artificially cut at $R_t$.
The only parameter that determines the profile is the ratio of $R_t$ to $R_c$,
$c_p$.  Here we perform two simulations analogous to model 13 with $c_p = 3$
and $c_p = 6$.  The Plummer model with $c_p=6$ has
core density and half-mass radius comparable to those of the King model with
$W_0=4$.  We find that the Plummer models give $t_{ev}$ values only slightly
longer than that of the corresponding King model (by less than 20~\%), and
$\Sigma_{20}$ and the number counts of the Plummer models are very similar to
those of the King model.  Therefore, the findings in \S~\ref{sec:init} may also
apply to models with Plummer initial conditions with $c_p = 3$--6.

\section{SUMMARY}
\label{sec:summary}

Using Aarseth's Nbody6 codes, we have studied the dynamical evolution
of CYCs near the GC.  First, we confirm the results of KML that
clusters with a mass $\lsim 2 \times 10^4 \msun$ evaporate in
$\sim 10$~Myr, but find that the F-P calculations by KML
underestimated the lifetimes of CYCs by 5 to 30~\% in most
cases.  This discrepancy is due to the adoption by KML of values of $N_u$
which are too small,
and we find that $N_u$ of 250--300 would be appropriate although such a
large number would bring the effctive mass of the largest mass bin down
to a value at which we would lose some of the effects associated with
the presence of the most massive stars.  Without stellar
evolution, the above $t_{ev}$ discrepancy would be more considerable,
i.e. the mass loss from stellar evolution starting slightly after $t=2$~Myr
significantly accelerates the evaporation of the clusters.  The volume density
profiles from F-P calculations are steeper than those from N-body simulations,
and especially, the F-P densities outside $R_t$ are significantly
underestimated.

By comparing the surface number density profiles and number counts in
2 mass bins at 2 radius bins between N-body simulations and {\it HST}/NICMOS
photometry of the Arches cluster, we find the following best-fit
initial conditions: $M= 2 \times 10^4 \msun$ and $\alpha=1.75$.
The relation between $M$ and $\alpha$ is constrained by the observed
number of stars heavier than $30 \msun$, 150.  Larger or smaller $\alpha$
values than 1.75 give less satisfactory number count fits to observations.
The presence of primordial binaries favors slightly smaller $\alpha$ values
than 1.75.
The fit of $\Sigma_{20}$ to observations indicates $R_t \simeq 1$~pc,
but $R_g$ is not well constrained.  The mean density of the bulge inside
the Arches cluster is suggested to be 5--$30 \times 10^2 \msun
{\rm pc}^{-3}$.  Also, we confirm the finding of KML that for clusters with
$\alpha \lsim 2$, $t_{ev}$ of a cluster with $m_l =0.1 \msun$ is
very close to that of a cluster with $m_l=1 \msun$.  However,
the number count plots seem to more support $m_l = 1 \msun$.
These findings are in agreement with the arguments by \markcite{M93}Morris
(1993) that the non-standard star formation environment near the GC may
lead to an IMF skewed toward relatively massive stars and having
an elevated lower mass cutoff.

The global evolution and $t_{ev}$ of CYCs are nearly independent of
the initial mass segregation and the choice of initial density profile.
Clusters with initial mass segregation and reverse segregation (more
heavy stars in the outer region) show very similar evolution except
in the very early phase ($< 0.5$~Myr).  Plummer initial models with
$c_p=3$ and 6 exhibit indistinguishable evolution from a King initial model.
Therefore, our findings regarding the initial conditions of the Arches cluster
appear to be insensitive to the initial mass distribution.

As discussed in KML, the effect of the gas left over from cluster
formation may be important to the early dynamical evolution of CYCs
because the remnant gas is thought to be blown away from the cluster
in the early phases by strong stellar winds or supernova explosions,
and such abrupt disappearance may significantly change the potential
of the cluster in a short amount of time.  Consequently, the role of
the remnant gas in early cluster evolution will be the subject of
a future study using N-body simulations.

\acknowledgements
S.S.K. is deeply grateful to Sverre Aarseth for generously providing us
with his Nbody6 codes and for his kind help with the codes.
S.S.K. also appreciates Pavel Kroupa, Kap Soo Oh, Simon Portegies Zwart,
and Koji Takahashi for helpful discussions.
We thank the referee for valuable suggestions and comments.
This work was partially supported by a NASA grant to UCLA, and H.M.L.
acknowledges the support from the KOSEF through grant 1999-2-113-001-5.


\clearpage

\begin{deluxetable}{cccccccccc}
\tablecolumns{10}
\tablewidth{0pt}
\tablecaption{Comparison of $t_{ev}$ between N-body \& F-P Calculations
\label{table:nbody2fp}}
\tablehead{
\colhead{} &
\colhead{} &
\colhead{$M$} &
\colhead{} &
\colhead{$m_l$} &
\colhead{} &
\colhead{$R_g$} &
\multicolumn{2}{c}{$t_{ev}$ (Myr)} &
\colhead{$\Delta' t_{ev}$} \\ \cline{8-9}
\colhead{Model} &
\colhead{KML} &
\colhead{(${\rm M_\odot}$)} &
\colhead{$\alpha$} &
\colhead{(${\rm M_\odot}$)} &
\colhead{$N$} &
\colhead{(pc)} &
\colhead{N-body} &
\colhead{F-P} &
\colhead{(\%)}
}
\startdata
1  & 142     & $2 \times 10^4$ & 2.35 & 1.0 &  6270 &  30 & 5.5 & 4.0 & 27 \nl
3  & 115     & $2 \times 10^4$ & 1.50 & 0.1 &  5164 &  30 & 2.8 & 2.2 & 21 \nl
5  & 141     & $5 \times 10^3$ & 2.35 & 1.0 &  1567 &  30 & 2.4 & 2.2 &  8 \nl
8  & 113     & $2 \times 10^4$ & 1.50 & 1.0 &  1633 &  30 & 2.7 & 2.3 & 15 \nl
13 & \nodata & $2 \times 10^4$ & 1.75 & 1.0 &  2605 &  30 & 2.8 & 2.4 & 14 \nl
15 & \nodata & $2 \times 10^4$ & 1.75 & 0.1 & 12706 &  30 & 3.0 & 2.6 & 13 \nl
23 & 101     & $2 \times 10^4$ & 2.00 & 1.0 &  6270 &  30 & 3.6 & 2.7 & 25 \nl
43 & \nodata & $2 \times 10^4$ & 1.75 & 1.0 &  2605 &  10 & 2.2 & 0.9 & 59 \nl
44 & \nodata & $2 \times 10^4$ & 1.75 & 1.0 &  2605 & 100 & 4.5 & 3.8 & 16 \nl
\enddata
\tablecomments{F-P values are from KML except for models with $\alpha=1.75$,
which were additionally calculated for the present study.  The coefficient
for the speed of star removal, $\alpha_{esc}$, of 2 is adopted for all
F-P simulations here.  Column KML is for the model numbers from the paper KML.
$\Delta' t_{ev}$ is the fractional difference, $|$FP-Nbody$|$/Nbody.
}
\end{deluxetable}

\begin{deluxetable}{ccc}
\tablecolumns{3}
\tablewidth{0pt}
\tablecaption{Number Counts from the Observation of the Arches Cluster
\label{table:obscount}}
\tablehead{
\colhead{} &
\multicolumn{2}{c}{Mass Bin} \\ \cline{2-3}
\colhead{Radius Bin} &
\colhead{$8 \leq m/{\rm M_\odot} \leq 20$} &
\colhead{$20 \leq m/{\rm M_\odot}$}
}
\startdata
$0.4 \leq r/{\rm pc} \leq 0.8$ & 64 & 46 \nl
$0.2 \leq r/{\rm pc} \leq 0.4$ & 66 & 56 \nl
\enddata
\end{deluxetable}

\begin{deluxetable}{cccccc}
\tablecolumns{6}
\tablewidth{0pt}
\tablecaption{N-body Simulations for the Arches' Initial Conditions
\label{table:init}}
\tablehead{
\colhead{} &
\colhead{$M$} &
\colhead{} &
\colhead{$m_l$} &
\colhead{$R_g$} &
\colhead{$t_{ev}$} \\
\colhead{Model} &
\colhead{(${\rm M_\odot}$)} &
\colhead{$\alpha$} &
\colhead{(${\rm M_\odot}$)} &
\colhead{(pc)} &
\colhead{(Myr)}
}
\startdata
13 & $2.0 \times 10^4$ & 1.75 & 1.0 &  30 & 2.8 \nl
14 & $2.0 \times 10^4$ & 1.75 & 1.0 &  50 & 4.1 \nl
15 & $2.0 \times 10^4$ & 1.75 & 0.1 &  30 & 3.0 \nl
19 & $2.0 \times 10^4$ & 1.75 & 1.0 &  20 & 2.4 \nl
21 & $1.6 \times 10^4$ & 1.50 & 1.0 &  30 & 2.6 \nl
22 & $2.8 \times 10^4$ & 2.00 & 1.0 &  30 & 3.7 \nl
\enddata
\end{deluxetable}

\clearpage

\end{document}